\documentclass[usenatbib]{mn2e}
\usepackage[utf8]{inputenc}
\usepackage{natbib,epsfig,cite,mystyle,amsmath,amssymb,graphicx}
\usepackage[font=scriptsize]{caption}
\usepackage{array, xcolor}
\usepackage[colorlinks=true,linkcolor=blue,citecolor=blue]{hyperref}

\title{The origin of dispersion in DLA metallicities}
\author[I. Dvorkin, J. Silk, E. Vangioni, P. Petitjean, K. Olive]{Irina
Dvorkin$^{1}$\thanks{E-mail:
dvorkin@iap.fr}, Joseph Silk$^{1,2}$, Elisabeth Vangioni$^{1}$, Patrick
Petitjean$^{1}$, Keith A. Olive$^{3}$\\
$^{1}$Sorbonne Universités, UPMC Univ Paris 6 et CNRS, UMR 7095, Institut
d’Astrophysique de Paris, 98 bis bd Arago, 75014 Paris, France \\
$^{2}$Department of Physics and Astronomy, The Johns Hopkins University,
Baltimore, MD 21218, USA \\
$^{3}$William I. Fine Theoretical Physics Institute, School of Physics and
Astronomy, University of Minnesota, Minneapolis, MN 55455, USA
}

\begin{document}

\pagerange{\pageref{firstpage}--\pageref{lastpage}} \pubyear{2015}
\maketitle
\label{firstpage}

\begin{abstract}

Recent chemical abundance measurements of damped Ly$\alpha$ absorbers (DLAs)
revealed an intrinsic scatter in their metallicity of $\sim 0.5$ dex out to
$z\sim 5$. In order to explore the origin of this scatter, we build a
semi-analytic model which traces the chemical evolution of the interstellar
matter in small regions of the Universe with different mean density, from over-
to underdense regions. We show that the different histories of structure
formation in these regions,
namely halo abundance, mass and stellar content, is reflected in the chemical
properties of the protogalaxies, and in particular of DLAs. We calculate
mean metallicity-redshift relations and show that the metallicity dispersion
arising from this environmental effect amounts to $\sim 0.25$ dex and is an
important contributor to the observed overall intrinsic scatter.
\end{abstract}


\section{Introduction}

Damped Ly$\alpha$ absorbers (DLA), defined as quasar absorption systems with
neutral column densities $N_{HI}>2\times 10^{20}$ cm$^{-2}$
\citep{1986ApJS...61..249W}, dominate the neutral gas content of the Universe
in the redshift range $z=0-5$ and are likely the progenitors of
low redshift galaxies \citep{2005ARA&A..43..861W}.
The chemical properties of DLAs can be determined with great precision, and
provide a unique probe of the properties of cold neutral gas out of which stars
form at high redshifts.

Numerous surveys conducted over the past several years have
provided extensive samples of high-redshift absorbers
\citep[e.g.][]{2005MNRAS.363..479P,2005ApJ...635..123P,2012A&A...547L...1N}.
High resolution observations of nearly $250$ of these systems established a
statistically significant decrease of DLA metallicity with increasing redshift
\citep{2003ApJS..147..227P,2012ApJ...755...89R} and a large intrinsic dispersion
of $\sim 0.5$ dex out to $z\sim 5$. Interestingly, this dispersion is not caused
by observational uncertainties and does not appear to evolve with redshift.
\citet{2013ApJ...769...54N} suggested
that this effect could be partially explained by the existence of a fundamental
plane in the redshift-mass-metallicity space, so that the dispersion in
the metallicity-redshift relation is caused by the scatter in the masses of the
dark matter (DM) halos hosting the DLAs \citep[see
also][]{2006A&A...457...71L,2013MNRAS.430.2680M}. There could be,
however, additional
sources of intrinsic scatter related to the environment of the galaxy or the
distribution of neutral clouds within the halo.

DLAs have been extensively studied with hydrodynamic simulations
\citep[e.g.][]{2004MNRAS.348..421N,2008MNRAS.390.1349P,
2011MNRAS.418.1796F,2014MNRAS.445.2313B}, which have been able to
reproduce the
observed DLA abundance and metallicity range. While hydrodynamic simulations
are becoming extremely accurate and are now able to explain many of the
properties of high-redshift galaxies and their surroundings, their high
computational cost prohibits studies of the large volumes of the Universe
necessary for obtaining good statistics. Therefore the study of DLA metallicity
dispersion with numerical simulations is extremely challenging.
Notably, \citet{2012ApJ...748..121C} simulated one overdense and one underdense
region, whose properties bracketed many of the observed properties of DLAs.

On the other hand, the semi-analytic approach, successfully employed in the
study of galaxy evolution, enables one to reproduce large galaxy populations and
perform a comprehensive parameter study at the cost of making several
simplifying assumptions. \citet{2014MNRAS.441..939B} studied DLA
properties in the context of a detailed semi-analytic model of galaxy formation
and were able to reproduce several key properties of DLAs, including the column
density distribution and the evolution of metallicity with redshift. However, these earlier models have so far just provided mean chemical evolution trends with $z.$

In this work,
we build a computationally efficient semi-analytic merger tree model specifically
designed to study the dispersion in the metallicity-redshift relation and the
history of chemical enrichment of the Universe. We use the DM
merger tree building algorithm from the GALFORM code
\citep{2000MNRAS.319..168C} to construct regions with different histories of
structure formation. We then apply the chemical evolution model of
\citet{2004ApJ...617..693D,2006ApJ...647..773D} to explore the evolution of ISM
metallicities in these different environments.

This paper is organized as follows. In Section \ref{sec:model} we describe our
model, which uses analytically computed DM halo merger trees and an accurate
model of chemical evolution. In Section
\ref{sec:results} we show that this model produces a dispersion
in the observed DLA metallicities. We discuss our results in Section
\ref{sec:dis}.

\section{Model}
\label{sec:model}

Our goal is to calculate the dispersion in the mean mass fraction of collapsed
structures, escape velocity and cosmic star formation rate in different parts of the Universe. As a
first step, we construct merger trees of DM halos using the
GALFORM algorithm \citep{2000MNRAS.319..168C}. This algorithm relies on the
extended Press-Schechter theory, in
particular the conditional mass function (MF) of DM halos, and provides a list
of mergers, and the redshifts at which they occurred, that led to the formation of
a given halo. The modification by \citet{2008MNRAS.383..557P} renders the
algorithm
consistent with the Sheth-Tormen MF \citep{1999MNRAS.308..119S}. Due to the
shape of the matter power
spectrum in cold DM models, the
probability of a given halo with mass $M_0$ to merge with another halo with mass
$M$ diverges as
$M\rightarrow 0$, therefore we employ a cut-off mass $M_{res}$, which sets the
mass resolution of our calculation. We use $M_{res}=M_{min}/10$ where $M_{min}$
is the smallest halo mass that is able to form stars. In what
follows, we explore the range $M_{min}=10^7-10^9 M_{\odot}/h$, and discuss this
choice below. More details on the merger-tree building algorithm can be found
in \citet{2000MNRAS.319..168C} and \citet{2008MNRAS.383..557P}.

We take $V_{tot}=10^6$ (Mpc/h)$^3$ as the total comoving volume of our
calculation, having checked that for this volume, the halo MF produced with the
merger-tree algorithm converges to the Sheth-Tormen MF at all redshifts, and
populate it with DM halos as follows. First we divide
the total mass range $M\in (M_{res}-10^{15}M_{\odot})$ into logarithmically
equal bins
of size $\Delta \ln M=0.01$. We
then calculate the mean number of halos in each bin at $z=0$
using the Sheth-Tormen MF
and draw the actual number
$N(M)$ from a Poisson distribution with this mean. Finally, we sample
$N(M)$ halos with logarithmically equally distributed masses in the
corresponding bin. 

We build a merger tree for each halo at $z=0$ and follow its evolution
backwards in time up to
$z_{f}=15$, saving the output in $50$ redshift bins equally spaced in redshift.
In this manner, we obtain the distribution of DM halos in the
comoving volume $V_{tot}$ as a function of redshift. We further divide our
sample into $N=1000$ smaller \emph{regions} $\Delta V_i=V_{tot}/N=10^3$
(Mpc/h)$^3$ by assigning each merger tree to a random region $i$. This
is equivalent to smoothing the density field on the scale of $r_0\sim 10$ Mpc/h
and assuming there is no correlation between neighbouring regions. This value
was chosen because on scales larger than $r_0,$ galaxies are only weakly
clustered \citep[e.g.][]{2001MNRAS.328...64N,2005ApJ...630....1Z}, and we expect
no significant dispersion in the properties of different
regions in the universe. On the other hand, for scales much smaller than $r_0$
there could be flow of matter between different regions so that they cannot be
treated as closed boxes. In other words, here we explore the effect of dispersion on the scale of massive clusters and deep voids; further contribution is expected from galactic scales, as we briefly discuss below. 
For comparison, in the simulations of
\citet{2012ApJ...748..121C} the overdense and underdense regions had a size of
roughly $20$ and $30$ Mpc/h on the side, respectively \citep[see also][]{2013MNRAS.428..540R}.

We then
calculate the mean mass fraction in collapsed structures $f_{coll}$, the mean
escape velocity and the mean star formation rate (SFR) in each region $\Delta
V_i$ as explained below. While the mean MF and $f_{coll}$ in our
total volume $V_{tot}$ coincide with the expectations from the Sheth-Thormen
MF at all redshifts, their values differ in each $\Delta V_i$ region. Those
regions that host a group or a cluster at low redshift have a higher
concentration of structures already present at higher redshift, whereas present
day voids are relatively empty at early times. This
small-scale inhomogeneity creates the observed dispersion
in the different observables.

The mean mass fraction in collapsed structures $f_{coll}$ is calculated by
summing the masses of all the halos above $M_{min}$ in the given region and
dividing by $\Delta V_i$ and the mean matter density $\rho_m$: 
\begin{equation}
 f_{coll,i}=\frac{\sum_{M_j>M_{min}} M_j}{\Delta V_i}\frac{1}{\rho_m}\:. 
\label{eq:fcoll}
\end{equation}
The dispersion
in $f_{coll}$ among the different regions grows with redshift for all values of
$M_{min}$, reaching $0.35$
dex for $M_{min}=10^9 M_{\odot}/h$ at $z=10$ and $0.25$ dex at $z=5$.
Furthermore, the dispersion grows with $M_{min}$, since if only large structures
are allowed to form, some regions remain practically devoid of structures. In
the case of $M_{min}=10^8 M_{\odot}/h$ and $M_{min}=10^7 M_{\odot}/h$ the
dispersion is about $0.25$ dex and $0.18$ dex at $z=10$, respectively.

The mean square of the escape velocity $v^2_{esc}(z)$ within each region is
calculated as the
mass-weighted average of $2GM/R$ for
each halo mass $M$, where $R$ is the virial radius enclosing a volume with mean
density $200$ times the critical density of the Universe.
Contrary to the case of the collapse fraction, the dispersion in $v^2_{esc}$ is
slightly larger for smaller values of $M_{min}$, being on average $0.2$ dex for
$M_{min}=10^7 M_{\odot}/h$. The reason for this is that models with
low $M_{min}$ exhibit a larger range of masses in each
region, so that while the total mass in two representative regions might be the
same, their mean escape velocities could differ substantially. In the case of
high $M_{min}$, however, the mass spectrum of different regions is much more
uniform and the main difference is in the total mass of collapsed structures,
reflected in $f_{coll}$. Also contrary to the case of $f_{coll}$ the dispersion
in $v^2_{esc}$ grows with decreasing redshift.
The mean escape velocity is used in our
chemical evolution code to calculate the outflow rate as explained below.

Our next step is to calculate the star formation rate (SFR) in each halo
and the mean cosmic SFR in each region $\Delta V_i$.
We use the results of \citet{2013ApJ...770...57B} (hereafter B13) to
calculate the SFR as a function of halo mass $M$ and redshift $z$ in the range $0<z<8$ and $9<\log_{10} (M/M_{\odot}) < 15$. We extrapolate these results to lower masses and higher redshifts with
$SFR\propto  (M/M_{\odot})^{\beta} 10^{\alpha z}$  where $\beta=2.1$ was obtained from a fit to the results of B13 in the range  $5<z<8$ and $9<\log_{10} (M/M_{\odot}) < 11$ and $\alpha=-0.1$ was obtained from the observed cosmic SFR at $z\sim 10$
\citep{2014ApJ...795..126B, 2014ApJ...786..108O}. We note that this extrapolation is purely phenomenological and IMF-independent.
We calculate
the SFR for each halo and then compute the cosmic SFR
density $\psi_i(t)$ as a function of redshift in each region $\Delta V_i$:
\begin{equation}
 \psi_i(t) = \frac{\sum_j SFR(M_j,t)}{\Delta V_i}\:.
\label{eq:sfr}
\end{equation}
We note that by using the fit from B13 (which refers to the universal SFR
density) in all of our calculations, in particular the small-scale regions, we
implicitly assume that the star formation efficiency depends only on the DM halo
mass and redshift. This approximation is justified by the fact that we compute
mean values over sufficiently large regions, which we treat as closed
boxes. The resulting dispersion in the SFR between different regions is roughly
$0.4$ dex, dropping significantly below $z\simeq 2.5$.

Having obtained the rate of growth of structure and the SFR density in each
region we proceed to the calculation of the chemical enrichment of the
interstellar matter (ISM). We use the chemical evolution model developed in
\citet{2004ApJ...617..693D,2006ApJ...647..773D} and \citet{2009MNRAS.398.1782R} which follows the exchange of
mass between the gas
within and outside of collapsed structures, the SFR at each redshift and the
rate of metal production in stars. 

The initial gas content of galaxies is taken
to be equal to the cosmic mean
$f_{baryon}=\Omega_b/\Omega_m$. The model then follows two gas reservoirs, the
intergalactic matter (IGM) and the ISM. Matter flows from IGM to ISM as the
galaxies form, where baryons are assumed to follow DM, so that the mean baryon
accretion rate in each region is given by $a_b(t)=\Omega_bdf_{coll}/dt$
and $f_{coll}$ is calculated as in eq. (\ref{eq:fcoll}).
Once inside galaxies,
baryons form stars according to the rate $\psi(t)$ calculated above (see eq.
(\ref{eq:sfr})). We assume a Salpeter initial mass function (IMF) $\Phi(m)$ for
$m_{inf}\leq m \leq m_{sup}$ with $m_{inf}=0.1 M_{\odot}$ and $m_{sup}=100
M_{\odot}$.

Baryons can flow from structures back to the IGM due to galactic winds or
feedback from SNe or AGN. In this work we consider outflows powered by stellar
explosions with the rate given by:
\begin{equation}
 o(t)=\frac{2\epsilon}{v^2_{esc}(t)}\int dm \Phi(m) \Psi(t-\tau(m)) E_{kin}(m)
\end{equation}
where $\tau(m)$ is the lifetime of a star with mass $m$, $E_{kin}$ is the
kinetic energy released when this star dies, $v_{esc}$ is the mean escape
velocity from structures and $\epsilon=0.0006$ is the fraction of
kinetic energy that powers the outflow. This value was chosen as in
\citet{2006ApJ...647..773D} through comparison of the model predictions for
the IGM component and the observed oxygen and carbon abundances in the
Ly$\alpha$ forest \citep[e.g.][]{2004ApJ...606...92S,2005AJ....130.1996S,2008ApJ...689..851A}. We
verified that small deviations from this fiducial value do not have a
significant effect on the resulting ISM metallicity.
The contribution of this term strongly
depends on $M_{min}$, since small halos have smaller escape
velocities and as a result are more easily disrupted. Consequently, models with
lower $M_{min}$ will be able to retain fewer baryons inside the galaxies and
will have lower ISM metallicities, as shown below.

To sum up, the evolution of the baryonic mass in the IGM and ISM is given by
$dM_{IGM}/dt=-a_b(t)+o(t)$
and $dM_{ISM}/dt=[-\psi(t)+e(t)]+[a_b(t)-o(t)]$
where $e(t)$ is the rate at which stellar mass is returned to the ISM by mass
loss or stellar deaths. These equations are solved for each region separately,
so that no matter flow is allowed between regions. This approximation is
justified by our choice of the region size.

The chemical evolution of the IGM and ISM is calculated as described in
\citet{2004ApJ...617..693D}. In particular, we do not employ
the instantaneous recycling
approximation but calculate the rate at which gas is returned to the ISM
including the effect of stellar lifetimes and computing stellar yields for each
element and for different stellar mass ranges.
Further details on the chemical evolution model can be found in
\citet{2004ApJ...617..693D,2006ApJ...647..773D} and \citet{2015MNRAS.447.2575V}.

In the next Section we show results for the metallicity of the ISM relative to
the solar value: $[M/H]=\log_{10}(M/H)-\log_{10}(M/H)_{\odot} $ and compare it
to DLA data.

\section{Results}
\label{sec:results}

\begin{figure}
\centering
\epsfig{file=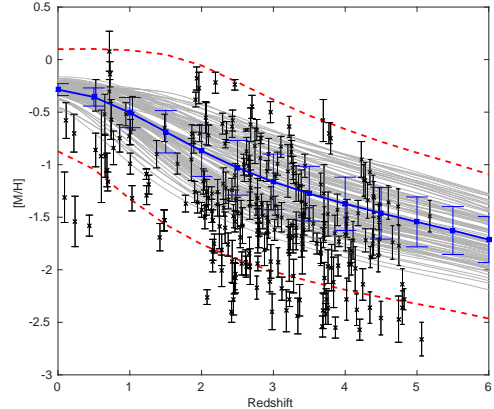, height=6cm}
\caption{Log of the metallicity abundance relative to the solar value for
$100$ regions, each
with a volume of $\Delta V_i=10^3$ Mpc$^3$/h$^3$ (thin grey lines) and their
mean (thick blue line) for $M_{min}=10^8 M_{\odot}/h$. The dispersion among the
realizations and their mean is represented by the blue points with error bars.
The mean dispersion in the redshift range $z=2-5$ is $0.25$ dex. Red dashed
lines represent the estimated upper and lower limits when the mass-metallicity
relation within each region is also considered (see text).
Black points represent data from \citet{2012ApJ...755...89R}.}
\label{fig:m1e8_Z}
\end{figure}

We can now test our model against observational data, using DLAs as probes of
the ISM at high redshift. For our fiducial model we
take $M_{min}=10^8 M_{\odot}/h$. We used $100$ regions,
each with $\Delta V_i=10^3$ (Mpc/h)$^3$, where both the mean and the
variance (calculated in redshift bins of $\Delta z=0.5$) over the whole ensemble
converge.

In Figure \ref{fig:m1e8_Z}, we show the evolution of the metal abundance relative
to the solar value for our fiducial model for
$100$ regions (grey lines) and compare them to observations of
DLAs from \citet{2012ApJ...755...89R}. The thick blue line shows the mean of all
the regions and the error bars represent their variance. It can be seen that
the dispersion predicted by the model ($0.25$ dex in the redshift range of the
observations) is significant, but smaller than the
observed dispersion in the metallicity-redshift relation ($0.5$ dex). 
We can thus conclude that the dispersion in the fraction of collapsed
structures, escape velocity and
SFR in each region $\Delta V_i$ contributes to the
dispersion in the metallicity-redshift relation of the DLAs, but is not its only
source. We note that the
dispersion predicted by our model is a lower limit on the actual value, since
the observed metallicity of a given DLA system might also depend on the mass of
the host DM halo. Indeed, \citet{2013ApJ...769...54N} showed that by applying a
correction following from
the existence of a global mass-metallicity-redshift plane, the scatter in the
metallicity-redshift relation
reduces from $0.5$ dex to $0.38$ dex. In addition, the potential metallicity gradients of DLAs can also contribute to the dispersion \citep{2005ApJ...629L..25C,2014MNRAS.445..225C}. Keeping in mind that the
observational
uncertainty for each measurement is about $0.12$ dex it is tempting to
suggest that the combination of a mass-metallicity-redshift plane and the
effect explored here, which stems mainly from the different formation epochs of
structures in over- and underdense regions, might
explain the observed dispersion. We stress that although we expect a correlation
between these two effects, they are somewhat distinct. Indeed, the
dispersion in the mean $\log M_{halo}$ \emph{between} regions is below $0.01$
dex for $z<5$.

While the modeling of the full effect is beyond the scope of the present
paper, we tried to estimate it using the observed mass-metallicity relation from
\citet{2008A&A...488..463M} and the dispersion in
stellar masses within each region using our merger tree, assuming for
simplicity a constant stellar mass to halo mass ratio. Since the
mass-metallicity relation is redshift related, and the parameters for high
redshifts are somewhat uncertain \citep[see also][]{2013ApJ...771L..19Z}, we
used the relation for $z=3.5$ for the whole
redshift range shown on Figure \ref{fig:m1e8_Z}. Adding the resulting
dispersion to the metallicities calculated with our model (grey lines) produces
the upper and lower limits shown as the red dashed lines on Figure
\ref{fig:m1e8_Z}. It can be seen that this very rough estimate can indeed
explain the whole obserbed range of DLA metallicities, although we caution that
a more self-consistent treatment is needed, which we leave to future work.

\begin{figure}
\centering
\epsfig{file=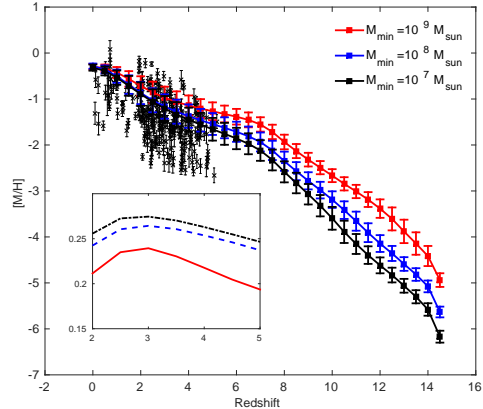, height=6cm}
\caption{Log of the metallicity abundance relative to the solar value,
averaged over $100$ regions. Error bars show the dispersion for each model with
$M_{min}=10^7,10^8$ and $10^9M_{\odot}/h$ (from bottom to top). Black points are measurements from
\citet{2012ApJ...755...89R}. The inset shows the dispersion (in dex) for
$M_{min}=10^7,10^8$ and $10^9M_{\odot}/h$ (from top to bottom).}
\label{fig:comp_mass}
\end{figure}

The choice of the minimal halo mass that is able to form stars affects both the
normalization and the dispersion of the predicted metallicity,
as can be seen in Figure \ref{fig:comp_mass} where we show model results for
$M_{min}=10^7,10^8$ and
$10^9M_{\odot}/h$ (from bottom to top). In the
case of high $M_{min}$ galaxies start forming later, but they are on average
more massive and have higher escape velocities. As a consequence, the outflows
are weaker and a larger fraction of metals produced in stars stay in the ISM as
compared to the case of low $M_{min}$. It is interesting to note that the
explored range $M_{min}=10^7-10^9 M_{\odot}/h$ matches the mean of the
observations. 
We note that the numerical
simulations of \citet{2004MNRAS.348..421N} show that the halo mass scale below
which no DLAs exist is slightly above $10^8 M_{\odot}$ at $z=3-4$, whereas
\citet{2008MNRAS.390.1349P} find that the occurrence rate of DLAs drops sharply
below $M\sim 10^{9}M_{\odot}$ \citep[although see][]{2015arXiv150402480W}. Our results are also consistent with these
limits.

The inset in Figure \ref{fig:comp_mass} shows the dispersion in metallicity
among $100$ regions as a function of redshift.
It can be seen that the dispersion is larger for smaller $M_{min}$. In
fact, there are two competing effects: in models with larger $M_{min}$ the
dispersion in $f_{coll}$ is larger, but the dispersion in $v^2_{esc}$ is
smaller. While $f_{coll}$ determines the overall number of galaxies in each
region, $v^2_{esc}$ influences the outflow, and is particularly important in
low-mass halos which could thereby lose a substantial fraction of their newly
acquired metals.

\section{Discussion}
\label{sec:dis}

In this paper, we have used the chemical evolution model of
\citet{2004ApJ...617..693D,2006ApJ...647..773D}, supplemented by a merger-tree
description of structure formation and an
observation-based estimate of the SFR, to study the metal enrichment process of
the ISM. Our models include the full range of galaxy formation physics required to study the chemical evolution of DLAs, including gas accretion, star formation rates, star formation histories, gas outflows and mergers.
We reached two important conclusions:

(i) The dispersion in structure formation rates
in different parts of the Universe and the accompanying variance in the stellar
content of protogalaxies produces a dispersion in the metallicity-redshift
relation of DLAs of roughly $0.25$ dex. Furthermore, this dispersion depends on
the
assumed minimal mass of halos that are able to form stars $M_{min}$ through the
dispersion in the structure formation histories: models with smaller $M_{min}$
produce larger dispersion. 

(ii) The range $M_{min}=10^7-10^9M_{\odot}/h$ provides a good
description of the data with little variance between these models. This mass
range corresponds to halo masses with virial temperatures
above $T \gtrsim 10^4$ for which neutral gas cooling is inefficient.

The dispersion resulting from our model is smaller than the observed
value of $\sim 0.5$ dex, but is clearly an important ingredient in the overall
effect. Another potentially important source of intrinsic scatter stems from the
difference in
the masses of DM halos hosting the DLAs, for example through the global
mass-metallicity-redshift relation discovered by \citet{2013ApJ...769...54N}.
Specifically, even within each over- or underdense region, there is a certain
spectrum of halo masses with different metal enrichment histories. According to
our very rough esimate, the combination of these two effects can explain the
observed dispersion, but a more rigorous treatment is needed, which takes into
account their possible correlations.
Another
important effect is related to the structure of the galactic halos giving rise to the DLAs, which may also contribute to the observed metallicity dispersion.

Although the conclusions outlined above are relatively robust, our model
involves several approximations which can be readily improved.
One such assumption is the SFR as a function of halo mass and redshift, in
particular our extrapolation of the results of B13 to high redshifts and low
masses, down to $M=10^7 M_{\odot}$. It would be beneficial to explore
more accurate and physically motivated extrapolation schemes, for example using
full semi-analytic models of galaxy evolution.

Galactic outflows obviously play an important role in galaxy evolution, and in
particular in the process of metal enrichment of the ISM. We have partly
addressed this issue by considering different values of $M_{min}$ which
resulted in different mean escape velocities and outflow rates. This
effect is crucial in setting the correct metallicity normalization, and we have
showed that models with lower $M_{min}$ have lower mean escape velocities and
retain less metals in the ISM. Clearly, a more comprehensive study of the
effects of different feedback mechanisms on the normalization and dispersion in
DLA metallicities is warranted, and we leave this to future work.

Finally, we plan to extend the model presented here to a full description of
chemical enrichment on a halo-to-halo basis. This will allow us to address the
issue of a global mass-metallicity-redshift relation of
DLAs, as well as the effect of the impact parameter
of quasar sightlines, thereby accounting for the different sources of
metallicity dispersion. 

\section*{Acknowledgments}                                                      
We thank the \textsc{galform} team for making the code
publicly available. The work of ID and JS was supported by the ERC Project No. 267117 (DARK) hosted by Universit\'{e} Pierre et Marie Curie (UPMC) - Paris 6, PI J. Silk. JS acknowledges the support of the JHU by NSF grant OIA-1124403. The work of KAO was supported in part by DOE grant DE-SC0011842 at the University of Minnesota. This work has been carried out at the ILP LABEX (under reference ANR-10-LABX-63) supported by French state funds managed by the ANR
within the Investissements d'Avenir programme under reference ANR-11-IDEX-0004-02, and was also sponsored by the French Agence Nationale pour la Recherche (ANR) via the grant VACOUL (ANR-2010-Blan-0510-01).

\bibliographystyle{mn2e}
\bibliography{dla}

\end{document}